\documentclass{aa}  

\usepackage{graphicx}
\usepackage{txfonts}
\usepackage{xcolor}
\usepackage{ulem}
\usepackage{hyperref}
\usepackage{orcidlink}
\hypersetup{
  colorlinks,
  citecolor=blue,
  linkcolor=blue,
  urlcolor=blue}

\begin{document}

   \title{Type Ia supernova explosion models are inherently multidimensional}

   \author{R\"udiger Pakmor\inst{1}\thanks{rpakmor@mpa-garching.mpg.de}\orcidlink{0000-0003-3308-2420}
          \and
          Ivo R. Seitenzahl\inst{2}\orcidlink{0000-0002-5044-2988}
          \and
          Ashley J. Ruiter\inst{2,3}\orcidlink{0000-0002-4794-6835}
          \and
          Stuart A. Sim\inst{4}\orcidlink{0000-0002-9774-1192}
          \and
          Friedrich K. R\"opke\inst{5,6}\orcidlink{0000-0002-4460-0097}
          \and\newline
          Stefan Taubenberger\inst{1}\orcidlink{0000-0002-4265-1958}
          \and
          Rebekka Bieri\inst{7}\orcidlink{0000-0002-4554-4488}
          \and
          St\'ephane Blondin\inst{8,9}\orcidlink{0000-0002-9388-2932}
          }

   \institute{Max-Planck-Institut f\"{u}r Astrophysik, 
              Karl-Schwarzschild-Str. 1, D-85748, Garching, Germany
              \and
              School of Science, University of New South Wales, Australian Defence Force Academy, Northcott Drive, Canberra, 2600, ACT, Australia
              \and
              ARC Centre of Excellence for All Sky Astrophysics in 3 Dimensions (ASTRO 3D), Canberra, ACT 2611, Australia
              \and
              School of Mathematics and Physics, Queen’s University Belfast, Belfast BT7 1NN, United Kingdom
              \and
              Heidelberg Institute for Theoretical Studies, Schloss-Wolfsbrunnenweg 35, D-69118 Heidelberg, Germany
              \and
              Zentrum f\"{u}r Astronomie der Universit\"{a}t Heidelberg, Institut f\"{u}r Theoretische Astrophysik, Philosophenweg 12, D-69120 Heidelberg, Germany
              \and
              Institut f\"ur Astrophysik, Universit\"at Z\"urich, Winterthurerstrasse 190, 8057 Z\"urich, Switzerland
              \and
              Aix Marseille Univ, CNRS, CNES, LAM, Marseille, France
              \and
              European Southern Observatory, Karl-Schwarzschild-Straße 2, Garching, D-85748, Germany
    }

   \date{Received 16/02/2024; accepted 20/04/2024}

  \abstract{
   Theoretical and observational approaches to settling the important questions surrounding the progenitor systems and the explosion mechanism of normal Type Ia supernovae have thus far failed. With its unique capability to obtain continuous spectra through the near- and mid-infrared, JWST now offers completely new insights into Type Ia supernovae. In particular, observing them in the nebular phase allows us to directly see the central ejecta and thereby constrain the explosion mechanism.
   
   We aim to understand and quantify differences in the structure and composition of the central ejecta of various Type Ia supernova explosion models.
   
   We examined the currently most popular explosion scenarios using self-consistent multidimensional explosion simulations of delayed-detonation and pulsationally assisted, gravitationally confined delayed detonation Chandrasekhar-mass models and double-detonation sub-Chandrasekhar-mass and violent merger models. 

   We find that the distribution of radioactive and stable nickel in the final ejecta, both observable in nebular spectra, are significantly different between different explosion scenarios. Therefore, comparing synthetic nebular spectra with JWST observations should allow us to distinguish between explosion models.

   We show that the explosion ejecta are inherently multidimensional for all models, and the Chandrasekhar-mass explosions simulated in spherical symmetry in particular lead to a fundamentally unphysical ejecta structure.

   Moreover, we show that radioactive and stable nickel cover a significant range of densities at a fixed velocity of the homologously expanding ejecta. Any radiation transfer postprocessing has to take these variations into account to obtain faithful synthetic observables; this will likely require multidimensional radiation transport simulations.
   }

   \keywords{supernovae: general; Hydrodynamics; white dwarfs}

   \maketitle

\begin{figure*}
    \centering
    \includegraphics[width=\textwidth]{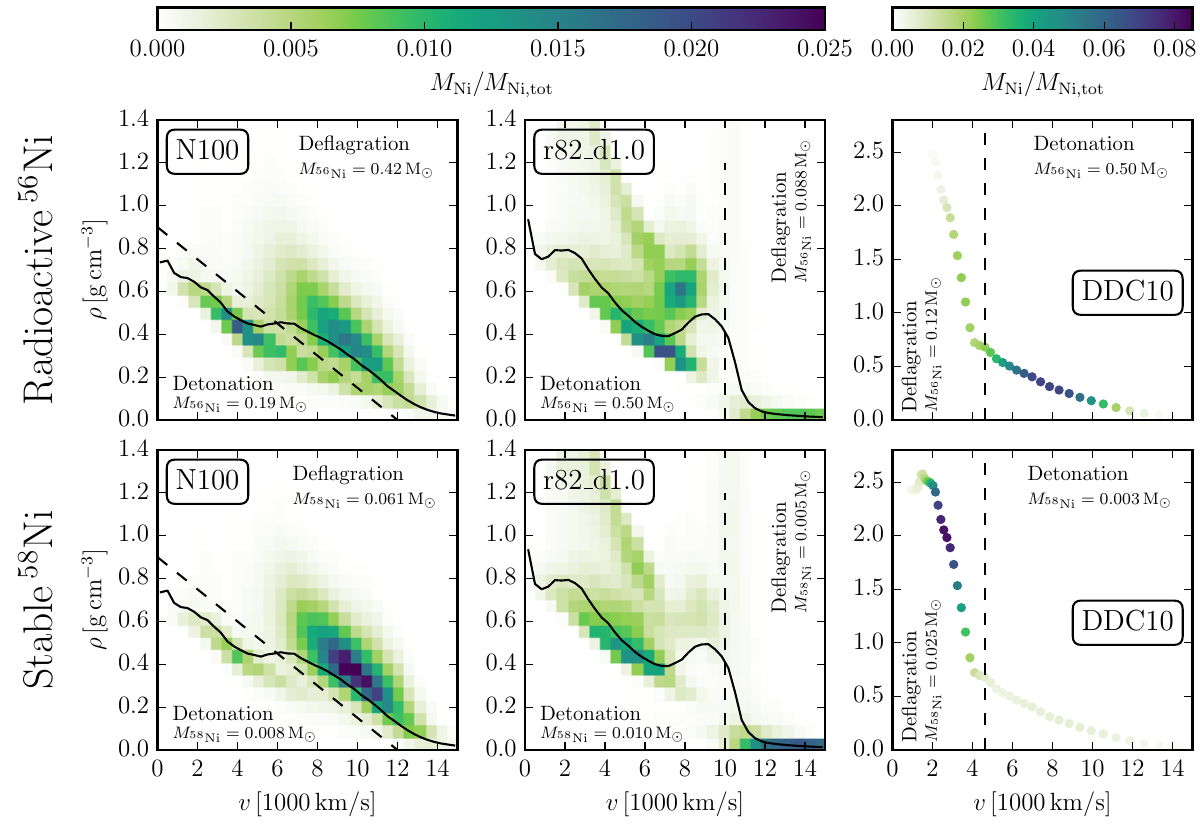}
    \caption{Distribution of radioactive $^{56}\mathrm{Ni}$ (top row) and stable $^{58}\mathrm{Ni}$ (bottom row) in density--velocity space $100\,\mathrm{s}$ after the explosion. Each histogram is normalised individually to the total mass of the type of nickel it shows. The columns show different Chandrasekhar-mass explosion models: a 3D delayed detonation \citep[\textbf{N100};][]{Seitenzahl2013b}, a 3D gravitationally confined detonation \citep[\textbf{r82.d1.0};][]{Lach2022}, and a 1D delayed detonation \citep[\textbf{DDC10};][]{Blondin2013}. The solid line in the left and middle panels shows the spherically averaged density profile of the total mass of the ejecta. The dashed lines indicate if nickel was synthesised in the deflagration or in the detonation. In the 3D models the deflagration ashes are located at higher velocities, i.e. further out in the ejecta, than the detonation ashes. In the 1D model this hierarchy is inverted. In the 3D models nickel is distributed over a wide range of densities at a fixed velocity, so spherical averaging 3D ejecta significantly changes their physical properties.}
    \label{fig:mch}
\end{figure*}

\section{Introduction}
Despite decades of observational and theoretical efforts, we still do not know the progenitor systems and the explosion mechanism of Type Ia supernovae. These questions are deeply connected but need to be answered individually. A detailed overview of the current state of the field is available in recent reviews \citep[e.g.][]{Ruiter2020,Liu2023}.

Spectra in the nebular phase of Type Ia supernovae more than $100\,\mathrm{d}$ after the explosion directly probe the centre of the ejecta, when they become optically thin \citep{RuizLapuente1992}. They, as well as supernova remnants \citep{Seitenzahl2019}, are thereby in principle well suited to distinguish explosion models.

The arrival of JWST and its ability to obtain continuous spectra of Type Ia supernovae in the nebular phase all the way through the near- and mid-infrared enables us not only to observe isolated nickel, cobalt, and iron lines in nebular spectra, but also to measure and compare their detailed line shapes \citep{Kwok2023, DerKacy2023, Blondin2023, Siebert2024, Kwok2024}. This new window into the inner ejecta of Type Ia supernova explosions may allow us to finally understand their explosion mechanism and solve the progenitor question.

This paper revisits existing explosion simulations of the most popular Chandrasekhar-mass and sub-Chandrasekhar-mass explosion scenarios currently in the running to explain normal Type Ia supernovae. We show that the ejecta of all these models are inherently multidimensional, and that both the explosion and very likely the radiation transport have to be simulated in more than one dimension to correctly interpret recent JWST observations.

We also show that the structure of the ejecta, and in particular the distribution of the stable and radioactive nickel in them, are characteristic of the different explosion scenarios. Both can in principle be directly probed with nebular spectra of Type Ia supernovae. Therefore, new and future JWST nebular spectroscopic observations of Type Ia supernovae will likely be very valuable for constraining and understanding their physics.

We start with a brief summary of the explosion physics of Type Ia supernovae in Section~\ref{sec:physics}. We then compare different explosion models of Chandrasekhar-mass white dwarfs in Section~\ref{sec:mch} and sub-Chandrasekhar-mass white dwarfs in Section~\ref{sec:submch}, and quantify the similarities and differences of the ejecta of different explosion scenarios. In Section~\ref{sec:implications} we discuss implications for obtaining synthetic observables and for the interpretation of observations. We conclude with a summary of our main results and a brief outlook in Section~\ref{sec:conclusions}.

\section{The physics of Type Ia supernova explosions}
\label{sec:physics}

To understand the connection between explosion mechanism and production of radioactive and stable nickel and the need for multidimensional models, we first took a step back and looked at what we know about Type Ia supernovae explosion physics.

There is a broad consensus that the progenitor systems of Type Ia supernovae are binary systems with at least one massive carbon-oxygen white dwarf. Since isolated white dwarfs are inert, we believe that interaction with a companion star leads to the explosion. Upon ignition, unstable nuclear burning disrupts the massive white dwarf and produces about a solar mass of ejecta with significant amounts of radioactive $^{56}$Ni, which decays via $^{56}$Co to $^{56}$Fe and powers the light curve of a Type Ia supernova \citep{Diehl2015, Churazov2015}. Here, we focused specifically on normal Type Ia supernovae that make up $\sim 70\%$ of all Type Ia supernovae and produce a typical amount of $0.6\pm 0.3\,\mathrm{M_\odot}$ of radioactive $^{56}$Ni \citep{Li2011,Taubenberger2017}.

To connect progenitor systems with observables of explosions and understand the production of radioactive and stable nickel, we needed to look at the explosion mechanism and the nuclear burning that powers the explosion. Unstable nuclear burning of degenerate matter proceeds as a thin burning front separating burning ashes and unburned fuel\footnote{Note, however, that at low very densities deflagration flames enter the distributed burning regime and cannot be considered as thin any longer \citep{Roepke2005}}. They come in two fundamentally distinct modes \citep[see e.g.][for a review]{Roepke2017}.

Deflagration flames propagate subsonically via thermal conduction. Therefore, the unburned fuel reacts to the approaching flame before it is burned. Moreover, the whole white dwarf responds to the energy release from the deflagration and expands on a timescale comparable to the timescale on which the deflagration burns the white dwarf.

Detonation fronts are in essence shock waves sustained by energy released from nuclear burning immediately behind the shock front. They propagate supersonically with respect to the sound speed of the fuel. A detonation burns the fuel before it can react to the approaching burning front. Once started, a detonation burns the whole white dwarf essentially instantaneously, faster than the sound crossing time of the white dwarf.

An important consequence is that fuel, as well as the white dwarf as a whole, can react to the approaching deflagration, but not to the approaching detonation. When a deflagration flame is ignited in a white dwarf, the energy release from nuclear burning causes the white dwarf to expand on a timescale comparable to the timescale on which the deflagration flame propagates through the white dwarf. Therefore, the density of a parcel of fuel when it is burned is typically significantly lower than its initial density was at the instant when the deflagration ignited. In contrast, a detonation burns all of the white dwarf's material essentially at the density it had at the moment the detonation formed.

The density of the fuel when it is burned by either a deflagration or a detonation is crucial because it principally determines the composition of the ashes. In particular, this includes the amount of neutronisation. Carbon-oxygen white dwarfs with solar metallicity are close to a $Y_\mathrm{e}=0.5$, that is, they have an equal number of protons and electrons. At densities $\rho \gtrsim 10^9\,\mathrm{g\,cm^{-3}}$ weak reactions become important enough to substantially neutronise the fuel when it is burned and decrease $Y_\mathrm{e}$ \citep{seitenzahl2009b}. This systematically changes the typical isotopes created in the nuclear burning.

Unfortunately, the ignition of white dwarfs is one of the unsolved questions in the puzzle of the explosion mechanism of Type Ia supernovae. Despite fundamental physical differences, different explosion models are not easily distinguished by their observables around peak brightness \citep{Roepke2012}, because the composition of the outer layers of the ejecta are typically similar between different explosion models and the observables are rather insensitive to most of the detailed structure of the inner, optically thick part of the ejecta.

Because neutronisation during nuclear burning strongly depends on density, the ratio between radioactive $^{56}$Ni ($Y_\mathrm{e}=0.5$) and stable $^{58}$Ni ($Y_\mathrm{e}<0.5$) becomes a sensitive probe of the burning conditions of the inner parts of the white dwarf. Moreover, because radioactive and stable nickel are both produced in substantial amounts in Type Ia supernovae they offer an important observational diagnostic \citep[see e.g.][]{Blondin2022}.

Radioactive $^{56}$Ni powers the light curve of Type Ia supernovae around maximum light via radioactive decay to $^{56}$Co and further to $^{56}$Fe. In the nebular phase, $100\,\mathrm{d}$ or later after the explosion, it has already completely decayed. Therefore, nickel detected in nebular spectra or supernova remnants is completely dominated by stable $^{58}$Ni that was synthesised directly in the explosion. In contrast, cobalt and iron in the nebular phase are dominated by decay products of $^{56}$Ni, and therefore trace the radioactive $^{56}$Ni created in the explosion. Observing nickel, cobalt, and iron in nebular spectra or remnants of type Ia supernovae thus directly informs us about the nucleosynthesis conditions in the centre of the explosion itself. It is therefore crucial to understand what to expect for different explosion scenarios to interpret observations.

Here we analysed the structure of the inner ejecta of multidimensional explosion simulations of the most popular Chandrasekhar-mass and sub-Chandrasekhar-mass explosion scenarios currently in the running to explain normal Type Ia supernovae \citep{Ruiter2020,Liu2023} with a focus on the distribution of radioactive and stable nickel in the ejecta.

\section{Chandrasekhar-mass models}
\label{sec:mch}

Chandrasekhar-mass explosion models start from a carbon-oxygen white dwarf that is accreting material from a companion star and is approaching the Chandrasekhar mass. The high pressure leads to nuclear reactions in its centre \citep{cameron1959a}, which slowly increase its temperature and drive large-scale convection \citep[e.g.][]{hoeflich2002a}. At some point, convection becomes insufficient to transport away the energy generated from nuclear burning \citep[e.g.][]{woosley2004a}. At this time, a local runaway leads to unstable nuclear burning and proceeds either as a deflagration or a detonation. This ignition is hard to model in detail, but probably occurs in a single spot slightly off-centre \citep{Zingale2011,Nonaka2012,Malone2014}. Importantly, at this point the central density of the white dwarf is typically higher than $10^9\,\mathrm{g/cm^3}$, and almost all of its mass sits above $10^7\,\mathrm{g/cm^3}$ \citep[e.g.][]{gasques2005a,lesaffre2006a,piersanti2022a}.

At the moment, we cannot make an ab initio statement about the nature of the initial burning. Ignition of either a deflagration or a detonation is in principle possible. If the initial ignition led to a detonation, all the mass of the white dwarf would burn at its density at ignition. Because most of the mass resides at a density larger than $10^7\,\mathrm{g/cm^3}$, the burning products are almost exclusively iron group elements, and significant production of intermediate-mass elements like silicon does not occur \citep{Arnett1969, Seitenzahl2017}. Such an explosion has never been observed even though it would be very bright.

Therefore, the initial ignition has to proceed as a deflagration. It initially burns at densities close to $10^9\,\mathrm{g/cm^3}$ in the centre of the Chandrasekhar-mass white dwarf and its energy release causes the white dwarf to expand. Because the deflagration starts burning at high densities, its ashes always contain a high fraction of stable nickel. The deflagration ashes rise buoyantly and drive turbulence in the whole white dwarf. For the most likely off-centre ignition \citep{Nonaka2012}, only part of the white dwarf is burned and ejected, typically leaving a bound remnant with a mass of ${\sim} 1\,\mathrm{M_\odot}$. The ejecta are dominated by iron group elements. Ejecta of such an explosion have been shown to provide a compelling match to the peculiar subclass of 02cx-like Type Ia supernovae \citep{Kromer2013,Magee2016,Kawabata2021,Lach2022Iax,Dutta2022,Singh2024}. For the energetically most optimistic central isotropic ignition the whole white dwarf does explode, but the ejecta do not at all resemble normal Type Ia supernovae \citep{Roepke2006,Roepke2007}.

However, it seems possible to start a detonation after the energy release from the initial deflagration has expanded the white dwarf significantly. When a significant fraction of the white dwarf has dropped below a density of $10^7\,\mathrm{g/cm^3}$, a detonation that burns the white dwarf essentially instantaneously at the density it has then, produces substantial amounts of intermediate mass elements. 

Two fundamentally different, physically plausible paths have been proposed to lead to a detonation. In the delayed detonation model \citep{Khokhlov1991}, strong turbulence in the white dwarf leads to a transition of the deflagration flame into a detonation. In the gravitationally confined detonation model \citep{Plewa2004}, a bubble of deflagration ashes leaves the surface of the white dwarf and drives a flow around it. The flow of deflagration ashes converges on the opposite side and starts a detonation via compression.

In both cases, the ashes of the deflagration have already risen substantially when the detonation starts, and the central part of white dwarf at this time is dominated by unburned material. The detonation then burns the rest of the white dwarf essentially instantaneously and produces some radioactive nickel with little stable nickel, because the lower density of the fuel avoids additional neutronisation. The detonation ashes then occupy the central parts of the final ejecta.
 
We show the structure of stable and radioactive nickel in the ejecta of Chandrasekhar-mass explosion models quantitatively in Figure~\ref{fig:mch}. The figure shows the mass distribution of radioactive and stable nickel in velocity--density space $100\,\mathrm{s}$ after the explosion when the ejecta are fully in homologous expansion. We focused on three representative, well-studied, Chandrasekhar-mass explosion models that all produce about $0.6\,\mathrm{M_\odot}$ of radioactive $^{56}$Ni, typical of a normal type Ia supernova. They are \textbf{N100}, a delayed detonation model \citep{Seitenzahl2013b}, where turbulence leads to a detonation after $0.9\mathrm{s}$, \textbf{82\_d1.0}, a gravitationally confined detonation \citep{Lach2022}, where a detonation forms from a converging shock on the surface of the white dwarf $6.4\mathrm{s}$ after the ignition of the initial deflagration, and \textbf{DDC10}, a 1D model for a delayed detonation \citep{Blondin2013} where the deflagration is artificially turned into a detonation $1.7\mathrm{s}$ after the ignition of the deflagration when the density at the flame drops below $2.7\times 10^7\,\mathrm{g/cm^3}$.

\begin{figure*}
    \centering
    \includegraphics[width=\textwidth]{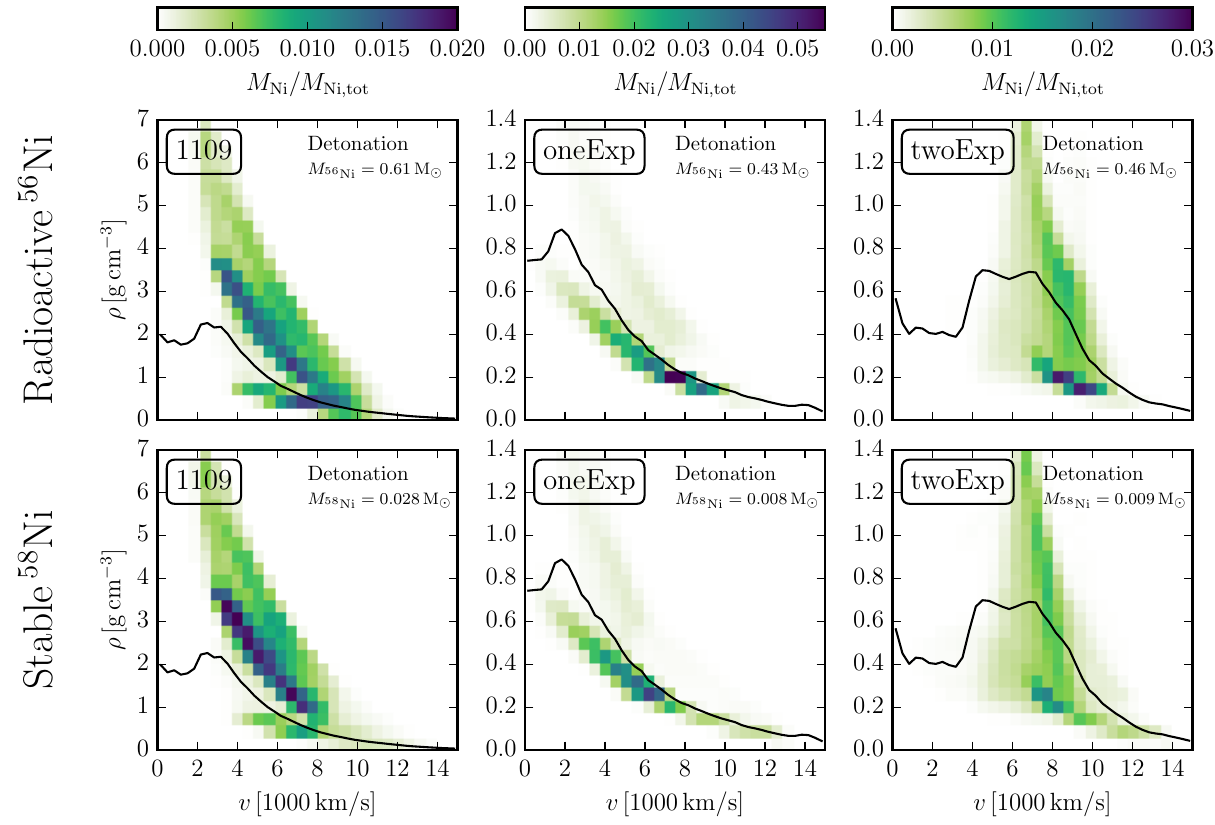}
    \caption{Distribution of radioactive $^{56}\mathrm{Ni}$ (top row) and stable $^{58}\mathrm{Ni}$ (bottom row) in density--velocity space $100\,\mathrm{s}$ after the explosion. Each histogram is normalised individually to the total mass of the type of nickel it shows. The columns show different sub-Chandrasekhar-mass merger explosion models:
    a violent merger \citep[\textbf{1109};][]{Pakmor2012} and a helium ignited merger in which only the primary white dwarf \citep[\textbf{oneExp};][]{Pakmor2022} explodes or both white dwarfs explode \citep[\textbf{twoExp};][]{Pakmor2022}. The solid line shows the spherically averaged density profile of the total mass of the ejecta. The distributions of radioactive and stable nickel in the ejecta are highly correlated. Only the oneExp model is close to spherical symmetry. In both the 1109 and twoExp models the secondary white dwarf explodes as well, compressing the ejecta of the primary white dwarf that contain primarily nickel. In the 1109 model the secondary explosion compresses the nickel to much higher densities than in the twoExp models because it happens with a delay of only $1\mathrm{s}$ compared to $4\mathrm{s}$ in the twoExp model.}
    \label{fig:submch}
\end{figure*}

\subsection{The separate location of radioactive and stable nickel in the ejecta}

In all Chandrasekhar-mass models shown in Figure~\ref{fig:mch}, radioactive and stable nickel are clearly separated. In the 3D models, the nickel created in the deflagration ends up at larger velocities than the nickel created in the detonation. This is a direct consequence of the buoyant rise in the deflagration ashes that drive Rayleigh--Taylor instabilities, leading to volume-filling turbulence in white dwarfs. The centre of the white dwarf is then filled with fuel when the detonation forms, and the central ejecta are made up exclusively of detonation ashes. 
The 1D model has a fundamentally different ejecta structure with an inverted hierarchy of detonation and deflagration ashes. This inversion is a direct result of the 1D nature of the model. It does not allow the model to describe the buoyant rise of deflagration ashes. Instead, they remain inside the detonation ashes that are created at larger radii after the deflagration has turned into a detonation. However, the buoyant rise of the deflagration ashes in 3D is unavoidable for all physically plausible ignition conditions. 

It has been suggested that a large-scale smooth initial deflagration front combined with an early ignition of a detonation could technically prevent the inversion of the ashes \citep{Leung2018, Kobayashi2020}. Crucially though, these conditions seem impossible to reach from physical ignition scenarios, because of buoyancy and Rayleigh--Taylor instabilities \citep{Zingale2011,Seitenzahl2013b}. Even implausibly strong magnetic fields cannot suppress these instabilities on large scales \citep{Hristov2021}. Moreover, the lack of turbulence in those models removes any physically plausible reason for the turbulent ignition of a detonation.

\subsection{Density variations in the ejecta at a fixed velocity}
 
As shown in Figure~\ref{fig:mch}, in all 3D Chandrasekhar-mass explosion models, nickel is spread over a significant range of densities at fixed velocity. The deflagration ashes in the N100 delayed detonation model are located in dense blobs, at a significantly higher density (up to factor of two) than the spherically averaged density of the total ejecta at the same velocity. Most of the detonation ashes in contrast sit at slightly lower densities than the spherical average of the ejecta, because of a tail to higher densities at low velocities. The 82\_d1.0 model shows a similar tail to nickel at higher densities that is caused by large-scale asymmetries in the white dwarf after the deflagration bubble has risen to the surface.
We note that microscopic mixing on unresolved scales will not change any of these conclusions, because the scales on which the deflagration ashes and detonation ashes are separated are well resolved in the simulations. The detonation ashes occupy comparable densities in the 3D models and the 1D model, but the 1D model has much higher densities in the inner deflagration ashes. A possible reason is that the deflagration ashes have to lift all of the unburned material of the white dwarf to expand. This challenge arises because, in the 1D model, the deflagration ashes cannot rise buoyantly and pass beyond the unburned material.

\section{Sub-Chandrasekhar-mass models}
\label{sec:submch}
 
The most important difference between explosion models of sub-Chandrasekhar white dwarfs and Chandrasekhar-mass white dwarfs is the density profile of the white dwarf at ignition \citep{Seitenzahl2017}. The central density of sub-Chandrasekhar-mass white dwarfs is substantially lower (a $1.2\,\mathrm{M_\odot}$ carbon-oxygen white dwarf has a central density of only $\sim 2\times 10^8\,\mathrm{g/cm^3}$). Deflagrations in these sub-Chandrasekhar-mass white dwarfs would cause the white dwarf to expand quickly, and only produce tiny amounts of $^{56}$Ni. Therefore, only immediate detonations are relevant for sub-Chandrasekhar-mass white dwarfs. The detonation at low densities avoids any significant neutronisation during nuclear burning. Thus, the amount of stable nickel produced in sub-Chandrasekhar-mass models is typically comparable to the amount created by the detonation in delayed detonation Chandrasekhar-mass models, but significantly lower than in their deflagration.

Sub-Chandrasekhar-mass explosion scenarios can be differentiated by their ignition mechanism. In the classic double detonation scenario, a sub-Chandrasekhar-mass carbon-oxygen white dwarf accretes helium until its helium shell reaches sufficient conditions in the densest regions to lead to unstable nuclear burning that starts a detonation. The detonation in the helium shell then wraps around the white dwarf, sending a shockwave into it from all directions. When the shockwave converges in its core (or potentially when it interacts with itself on the opposite site of the ignition point or when it first hits the carbon-oxygen core) it starts a carbon detonation that burns the whole white dwarf \citep{Bildsten2007, Fink2007, Fink2010, Shen2014b, Gronow2021b}.

If the companion is another carbon-oxygen white dwarf that donates helium from a thin shell, the helium detonation can form dynamically \citep{Guillochon2010, Pakmor2013, Pakmor2021, Pakmor2022}. This happens only when the secondary white dwarf is so close that it is about to be disrupted. At this time, the material in the accretion stream can be dense enough to be degenerate. Its impact on the surface of the primary white dwarf starts a detonation in the helium shell of the primary white dwarf like a hammer hitting an anvil. This scenario works for significantly smaller helium shells on the primary white dwarf, but for the purpose of this paper, it is essentially identical to the traditional double detonation scenario if the secondary white dwarf survives \citep{Boos2021,Shen2021,Pakmor2022}. However, the secondary white dwarf can probably ignite as well via a similar mechanism, where the impact of the ejecta of the exploding primary white dwarf starts a detonation in the helium shell of the secondary. In this case it also explodes, but with a delay of a few seconds, significantly changing the inner ejecta of the explosion \citep{Pakmor2022,Boos2024}.
Finally, if there is little to no helium present in a binary system of two carbon-oxygen white dwarfs, the disruption of the secondary white dwarf when the system merges and the impact this has on the primary white dwarf might directly lead to a carbon detonation on the surface of the primary white dwarf in the violent merger scenario \citep{Pakmor2010,Pakmor2011,Pakmor2012b}. In this case, both white dwarfs are burned by a detonation with only a short delay.

Figure~\ref{fig:submch} shows the mass distribution of radioactive and stable nickel in velocity--density space $100\,\mathrm{s}$ after the explosion for three typical sub-Chandrasekhar-mass explosion models. At this time the ejecta are fully in homologous expansion. 

The \textbf{1109} model is a violent merger of a $1.1\,\mathrm{M_\odot}$ and a $0.9\,\mathrm{M_\odot}$ white dwarf \citep{Pakmor2012}. The carbon detonation starts on the surface of the primary white dwarf when it is hit by the secondary white dwarf. It first burns the primary, then, with a delay of ${\sim}\, 1\mathrm{s}$, also the secondary white dwarf. Only the detonation that burns the primary white dwarf produces nickel; the central density of the secondary is too low to burn to iron group elements in the detonation in this case (but we note that the production of some nickel in the secondary would occur for an initially more massive secondary white dwarf). The ejecta of the secondary white dwarf expand into the central ejecta of the primary and compress them. Because of the small delay of only $\sim 1\mathrm{s}$, the inner ejecta of the explosion of the primary white dwarf, which are nickel-rich, are significantly compressed. Most nickel in the final ejecta ends up at densities significantly larger than the average density of the ejecta at the same velocity, because a significant amount of material of the ejecta of the secondary ends up at similar velocities. The final ejecta of 1109 at velocities smaller than $2000\,\mathrm{km/s}$ contain essentially no nickel and are made up purely by the ashes of the secondary white dwarf. 

The \textbf{oneExp} model is a double detonation explosion in a binary system of a $1.05\,\mathrm{M_\odot}$ and a $0.7\,\mathrm{M_\odot}$ white dwarf with thin helium shells of $0.03\,\mathrm{M_\odot}$ on each of them \citep{Pakmor2022}. Helium accretion on the primary white dwarf dynamically starts a helium detonation just a few orbits before the secondary would be disrupted and otherwise merge with the primary white dwarf. The helium detonation triggers a secondary carbon detonation in the core of the primary. In the oneExp model, the secondary white dwarf survives the explosion, by assumption (cf. twoExp below). The ejecta of the explosion are close to spherical, uniquely so among all 3D explosion models we consider here. The inner ejecta of the oneExp model are practically identical to classic double detonation models and simplified models of artificially centrally ignited sub-Chandrasekhar-mass carbon-oxygen white dwarfs \citep{Sim2010, Shen2018, Shen2021RT}.

The \textbf{twoExp} model starts out identical to the oneExp model, but after the ejecta of the explosion of the primary white dwarf hit the secondary, a similar double detonation mechanism is assumed to occur there as well: the impact of the ejecta of the primary detonates its helium shell and causes a carbon core detonation, which destroys the secondary \citep{Pakmor2022, Boos2024}. Similar to the 1109 violent merger model, the ejecta of the explosion of the secondary white dwarf do not produce any iron group elements and expand into the centre of the ejecta of the primary white dwarf, albeit with a much longer delay of ${\sim}\, 4\mathrm{s}$ (compared to ${\sim}\, 1\mathrm{s}$ for the violent merger model). They still significantly compress the nickel-rich inner ejecta of the explosion of the primary white dwarf, but much less so than in the 1109 model. They also dominate the ejecta out to larger velocities (about $5000\,\mathrm{km/s}$ compared to the $2000\,\mathrm{km/s}$ in 1109).

\section{Implications for Type Ia supernova nebular spectra and comparison to observations}
\label{sec:implications}

The result that the ejecta of Type Ia supernovae are asymmetric is not new, and has been discussed in detail already in the context of Chandrasekhar-mass delayed detonation models \citep{Gerardy2007,Kasen2009,Maeda2010} and violent mergers \citep{Bulla2016}. Our results emphasise that this is true in general for essentially all currently discussed explosion models of Type Ia supernovae. These asymmetries are fundamental to all 3D explosion models and faithfully simulating Type Ia supernova explosions crucially requires multidimensional simulations. In addition, we directly show that spherically averaging the ejecta of multidimensional explosion simulations before computing synthetic observables fundamentally changes the ejecta structure of the explosion and that 1D Chandrasekhar-mass explosions models are inherently unphysical. Therefore, extreme care should be taken when observational results are based on 1D explosion models \citep[e.g.][]{DerKacy2023}, or for 1D radiation transfer simulation of multidimensional explosion simulations.

The detonation ashes of all 3D Chandrasekhar-mass models and the oneExp model end up at similar densities in the final ejecta. Regions that have both higher density and large mass fractions of radioactive nickel are only present in the 1109 model (mass density much higher than the Chandrasekhar-mass models) and the twoExp model (factors of a few higher). In both sub-Chandrasekhar-mass models, the ejecta of the primary white dwarf are compressed to higher densities by the impact of the ejecta of the secondary white dwarf that explodes slightly later. The densities directly influence the ionisation states of nickel and iron, which are crucial for the formation and intensity of specific lines in the nebular spectra \citep{Floers2020,Blondin2023}. 
 
Different timings between the explosions of the primary and the secondary white dwarf are likely possible for merger models, at least in the whole range between $1\,\mathrm{s}$ (1109) and $4\,\mathrm{s}$ (twoExp). Nebular spectra covering a wide range of wavelengths through the infrared might allow us to infer the density of nickel and iron in the ejecta and therefore constrain properties of the progenitor system.

In sub-Chandrasekhar-mass models, the radioactive and stable nickel are co-located. Therefore, we naively expect similar line shapes for nickel and iron. In Chandrasekhar-mass models the radioactive and stable nickel are globally much less correlated. Still, the detonation ashes, which dominate the central ejecta for multidimensional models, contain about the same ratio of stable to radioactive nickel as in sub-Chandrasekhar-mass models. 
The deflagration ashes of Chandrasekhar-mass models, however, which contain a much higher ratio of stable to radioactive nickel, are located at much higher velocities in the ejecta. In the most extreme model, the gravitationally confined detonation model r82\_d1.0, the deflagration ashes sit at velocities larger than $10000\,\mathrm{km/s}$.
If the emission in nebular spectra of Chandrasekhar-mass models is dominated by the detonation ashes they might look very much like sub-Chandrasekhar-mass models. Detecting high-velocity stable Nickel that is fundamental to multidimensional Chandrasekhar-mass models, but seems impossible to get in sub-Chandrasekhar-mass models, could be one of the strongest possible links from observations to explosion scenario.

Crucially, this also means that the global ratio of stable to radioactive nickel in the ejecta that is currently often used to compare models and observations \citep{Maguire2018,Dhawan2018,Floers2020,Blondin2022,Liu2023}, might be a poor indicator for nebular spectra properties of Type Ia supernovae explosion models. Type Ia supernova remnants might be different in this aspect, because all radioactive nickel and cobalt has long decayed and their luminosity comes from the reverse shock \citep{Seitenzahl2019}. Nevertheless, we expect that for Chandrasekhar-mass explosions the nickel and iron lines in remnants will predominantly originate from different regions. Therefore, it is important to take the full multidimensional density and composition structure of the ejecta into account for any analysis of late-time Type Ia supernova observables.

\section{Conclusions}
\label{sec:conclusions}

We have shown that we really need 3D explosion simulations and likely 3D radiative transfer simulations to obtain faithful synthetic observables of the ejecta of Type Ia supernova explosion models.

We have focused on various models under consideration to explain normal Type Ia supernovae. Of all the models we included, only the double detonation scenario with a surviving companion has inner ejecta that are anywhere close to spherical symmetry.

We have shown that the central ejecta of 3D Chandrasekhar-mass delayed detonation explosion models are detonation ashes. The deflagration ashes have velocities higher than $5000\,\mathrm{km/s}$. In contrast, 1D Chandrasekhar-mass explosion models inherently lead to an unphysical inversion of the detonation and deflagration ashes in the ejecta (Figure~\ref{fig:mch}). In 3D Chandrasekhar-mass explosion models, both radioactive and stable nickel are distributed over a range of densities that varies by up to a factor of two in the ejecta at the same velocity.

In sub-Chandrasekhar-mass merger models in which the secondary white dwarf also explodes (violent merger or helium ignited merger), the nickel-rich ashes of the primary white dwarf can be substantially compressed by the ejecta of the secondary white dwarf. The details of the progenitor and explosion mechanism imprint themselves on the timing difference between the explosion of the primary white dwarf and the secondary white dwarf ($1\,\mathrm{s}$ for the violent merger, $4\,\mathrm{s}$ for the helium-ignited merger) as well as the magnitude of this compression. Since the nebular emission is highly sensitive to the range of densities that contribute to the emitting region of the ejecta, this could be used to learn about the progenitor systems of Type Ia supernovae in the future.

Recent and future JWST nebular spectra of Type Ia supernovae allow us to significantly improve our interpretation of observed Type Ia supernovae and match them with theoretical explosion models \citep{Kwok2023, DerKacy2023, Blondin2023, Kwok2024}. However, to do so we need faithful multidimensional models that simulate the explosion and obtain synthetic observables. Until we can compute nebular spectra and remnant models fully in 3D with all necessary physics involved, we will need to develop new, better approximations than spherically symmetric averaging \citep[e.g.][]{Blondin2023} and show that they represent full 3D models at least reasonably well.

\begin{acknowledgements}
AJR acknowledges support from the Australian Research Council under award number FT170100243. This research was supported in part by the National Science Foundation grant PHY-1748958 to the Kavli Institute for Theoretical Physics (KITP). ST acknowledges funding from the European Research Council (ERC) under the European Union’s Horizon 2020 research and innovation programme (LENSNOVA: grant agreement no. 771776).
The work of FKR is supported by the Klaus-Tschira Foundation, by the Deutsche Forschungsgemeinschaft (DFG, German
Research Foundation) --  537700965, and by the European Union (ERC, ExCEED, project number 1010962). Views and opinions expressed are however those of the author(s) only and do not necessarily reflect those of the European Union or the European Research Council Executive Agency. Neither the European Union nor the granting authority can be held responsible for them.
SAS acknowledges funding
from STFC grant ST/X00094X/1. This work was supported by the `Programme National de Physique Stellaire' (PNPS) of CNRS/INSU cofunded by CEA and CNES.
\end{acknowledgements}

\bibliographystyle{aa}

\end{document}